\documentclass[%
superscriptaddress,
 prl,
rsi,%
amsmath,amssymb,
reprint,%
floatfix,%
]{revtex4-1}

\usepackage{graphicx}
\usepackage{dcolumn}
\usepackage{xcolor}
\usepackage{float}

\usepackage{xspace}
\usepackage{amsmath} 
\newcommand{\ket}[1]{\ensuremath{|#1\rangle}\xspace}
\newcommand{\bra}[1]{\ensuremath{\langle #1|}\xspace}

\usepackage{multirow}
\usepackage{array}

\usepackage{lipsum}
\usepackage{subfigure}
\usepackage[justification=raggedright]{caption}

\begin{document}
\author{Enguerran Belles}
\affiliation{ 
Université de Lyon, Université Claude Bernard Lyon 1, CNRS, Institut Lumière Matière, UMR5306, F-69622 Villeurbanne, France
}%
\author{Franck Rabilloud}%
\affiliation{ 
Université de Lyon, Université Claude Bernard Lyon 1, CNRS, Institut Lumière Matière, UMR5306, F-69622 Villeurbanne, France
}%
\author{Alexander I. Kuleff}
\affiliation{ 
Theoretische Chemie, PCI, Universit{\"a}t Heidelberg, Im Neuenheimer Feld 229, D-69120 Heidelberg, Germany
}%
\author{Victor Despr\'e}%
\email{victor.despre@univ-lyon1.fr}
\affiliation{ 
Université de Lyon, Université Claude Bernard Lyon 1, CNRS, Institut Lumière Matière, UMR5306, F-69622 Villeurbanne, France
}%

\title{Size effect in correlation-driven charge migration in correlation bands of alkyne chains}

\date{\today}

\begin{abstract}
Correlation-driven charge migration initiated by inner-valence ionization leading to the population of the correlation bands of alkyne chains containing between 4 and 12 carbon atoms is explored through \textit{ab initio} simulations. Scaling laws are observed, both for the timescale of the charge migration and for the slope of the density of states of the correlation bands. Those can be used for predicting the relaxation time scale in much larger systems from the same molecular family and for finding promising candidates for the development of attochemistry scheme taking advantages of the specificity of the dynamics in correlation bands of molecules.
\end{abstract}

\maketitle

\section{Introduction}

Ultrafast science and technology have allowed the investigation of electron dynamics in molecules in their intrinsic time scale \cite{nisoli2017attosecond,lepine2014attosecond}. Studies were first focused on small systems for which fundamental mechanisms such as charge localization \cite{kling2006control}, instantaneous dipole moment \cite{neidel2013probing} or resonances \cite{haessler2009phase} have been observed at their fundamental quantum level. More recently, the experimental capabilities to tackle larger systems \cite{calegari2014ultrafast,marciniak2019electron} has created an effervescence around the possibility to develop an attochemistry. The main idea lying at the heart of attochemistry is to use electronic coherences to impact the reactivity of a molecular system. Indeed, electronic coherences lead to pure electron dynamics, like the extensively studied \cite{kuleff2010ultrafast,kuleff2016core,mignolet2014charge,sun2017nuclear,lara-astiaso2016decoherence,lara-astiaso2018attosecond,yuan2019ultrafast,folorunso2023attochemistry,haase2021electron} charge migration mechanism proposed by Cederbaum and coworkers \cite{cederbaum1999ultrafast,kuleff2014ultrafast}, that, if controlled, may make possible the preselection of reaction path. The realization of such a control scheme necessitates the existence of several prerequisites.

The first prerequisite is long-lasting electron coherences giving enough time for quantum control. Theoretical prediction have been made both in favor \cite{despre2015attosecond,despre2018charge,scheidegger2022search} and against \cite{vacher2015electron,jenkins2016nuclear,vacher2017electron,arnold2017electronic} the existence of long-lived electron coherences. While promising experimental observations have been made \cite{calegari2014ultrafast,kraus2015measurement,lara2018attosecond} it is only recently that the time of a long-lasting coherence was directly measured \cite{matselyukh2022decoherence}. This was done in silane (SiH$_4$), where an electron coherence of more than 10~fs, appearing as charge migration oscillations with a period of around 1.3~fs, has been observed. Moreover, after the initial loss of coherence, a revival appears around 50~fs. While it is usually considered that the loss of coherence is irreversible, this experiment showed that for small systems also revivals can occur (see also \cite{Jia2019timing}).

The second prerequisite is the ability of charge migration to influence the reactivity of a molecular system. Recently, such an impact has been experimentally observed for adenine \cite{maansson2021real}. A sub 3~fs delay in the stable dicationic signal has been detected in this XUV-pump IR-probe experiment. It was proposed \cite{despre2022correlation} that the dynamics responsible for this stabilization of the molecular cation is the correlation-driven charge migration taking place in a specific correlation regime, typical for inner-valence ionization of a molecule, in which the so-called breakdown of the molecular-orbital picture occurs \cite{cederbaum1986correlation}. The stabilization happens due to the characteristic charge-migration dynamics in which the shape of the initially created hole changes from a localized hole with $\sigma$ character to a delocalized hole with $\pi$ character, removing in this way the electron deficiency away from the bonds \cite{despre2022correlation}.

A consequence of the breakdown of the molecular orbital picture is the creation of a quasi-continuum of cationic states forming band-like structures starting just below the double ionization threshold of a molecule. As the breakdown is a consequence of electron correlation these structures have been called \textit{correlation bands} (CB) \cite{deleuze1996formation}. The dynamical consequence of populating these structures has been observed for the first time for a series of polycyclic aromatic hydrocarbons (PAH) \cite{herve2021ultrafast}. These XUV-pump IR-probe experiments were able to measure the relaxation time of the vibronic states in the correlation bands or, in other words, to obtain the time scale of the energy transfer between the electronic and nuclear degrees of freedom in this class of molecules. The time scale was observed to increase as a log-function with respect to the number of valence electrons of the molecular systems. Using a model based on an electron-phonon coupling calibrated with quantum electron-nuclear simulations \cite{marciniak2019electron}, a simple law has been derived containing 3 main parameters: the mean electron–phonon coupling strength, the mean phonon energy, and the slope of the density of cationic states (DOS) of the molecule as a function of energy. This relation permitted the interpretation of other previously published results \cite{belshaw2012observation,marciniak2015xuv,marciniak2018ultrafast}. As this initial non-adiabatic relaxation occurs through the coupling to specific vibrational modes, the follow-up thermalization, or redistribution of energy between the nuclear degrees of freedom, was also observed to follow a specific scaling law that depends on the number of valence electrons \cite{boyer2021ultrafast}.

These experimental observations and their theoretical explanations suggest that the correlation-driven charge migration in CB, being the initial response of the system to an inner-valence ionization, can have a direct impact on the molecular stability \cite{maansson2021real,despre2022correlation}. The correlation band of a molecule thus appears to be a promising playground for exploring the paradigm of attochemistry. 

With this paper, following the idea of developing a molecular design based on electron correlation \cite{despre2019size,folorunso2021molecular,mauger2022charge,chordiya2023photo}, we want to shed light on the behavior of the charge migration taking place after populating the correlation band of a series of molecules. For this purpose, we chose to study alkyne chains of increasing size: diacetylene (C$_4$H$_2$), the 1,3,5-hexatriyne (C$_6$H$_2$), the 1,3,5,7-octatetrayne (C$_8$H$_2$), the 1,3,5,7,9-decapentayne
(C$_{10}$H$_2$) and the 1,3,5,7,9,11-dodecahexayne (C$_{12}$H$_2$). This choice was made as it has been shown that the correlation-driven charge migration triggered by outer-valence ionization follows a scaling law with the number of valence electron \cite{despre2019size}. Furthermore, alkyne chains are used as building blocks for design of molecular-scale wires \cite{xiang2016molecular,harriman1996making,grosshenny1996towards,james2005molecular}. Here, we want to extract possible scaling law for the time scale of the correlation-driven charge migration in CB, as well as for the evolution of the DOS in the CB, in order to help in the development of model for the description of the follow-up vibronic relaxation in the correlation bands of carbon chains.

\section{Results and discussion}

The results presented here were obtained using the Algebraic Diagrammatic Construction (ADC) \cite{schirmer1982beyond,schirmer2018many} scheme to represent the Green's function at third order in perturbation theory in its non-Dyson formulation \cite{schirmer1998non}, that is nD-ADC(3). Third order means that all one-hole (1h) and two-holes--one-particle (2h1p) excitation are considered. The 1h excitation consists of the removal of an electron from an occupied orbital, while the 2h1p excitation consists of the removal of two electrons from occupied orbitals and the creation of an electron in a virtual one. ADC offers the crucial advantage to be able to obtain all cationic eigenstates of a molecular systems in a single calculation, which is primordial considering the high DOS in CB. The calculations were done with DZP basis and the molecules' geometry were optimized at the DFT B3LYP/DZP level, using Gaussian09 software \cite{frisch2009uranyl}.  

\begin{figure}[!ht]
  \centering
  \begin{subfigure}{}
  \includegraphics[scale=0.40]{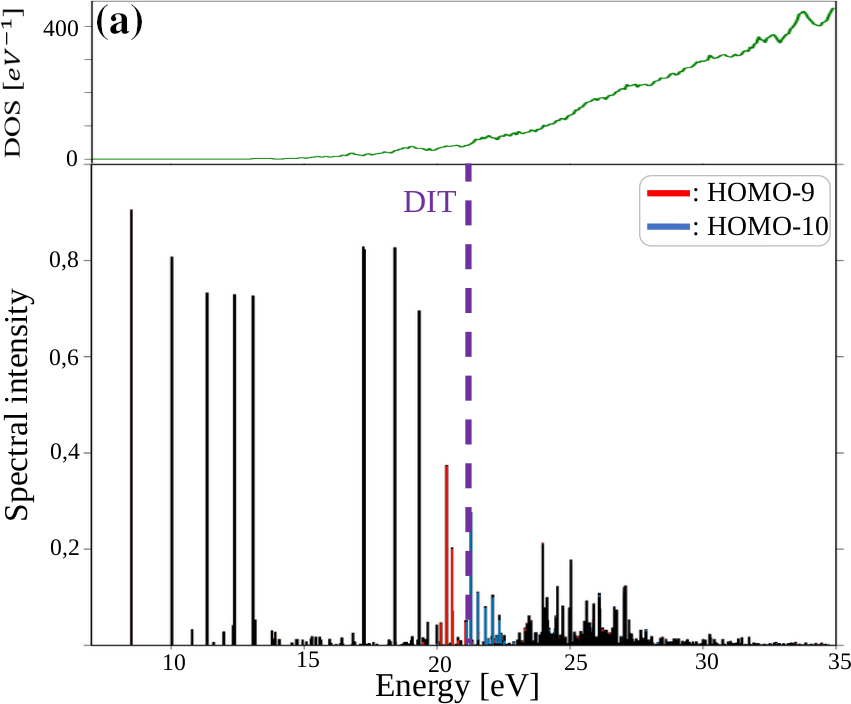}
   \label{fig:spec}
  \end{subfigure}
  \begin{subfigure}{}
    \includegraphics[scale=0.40]{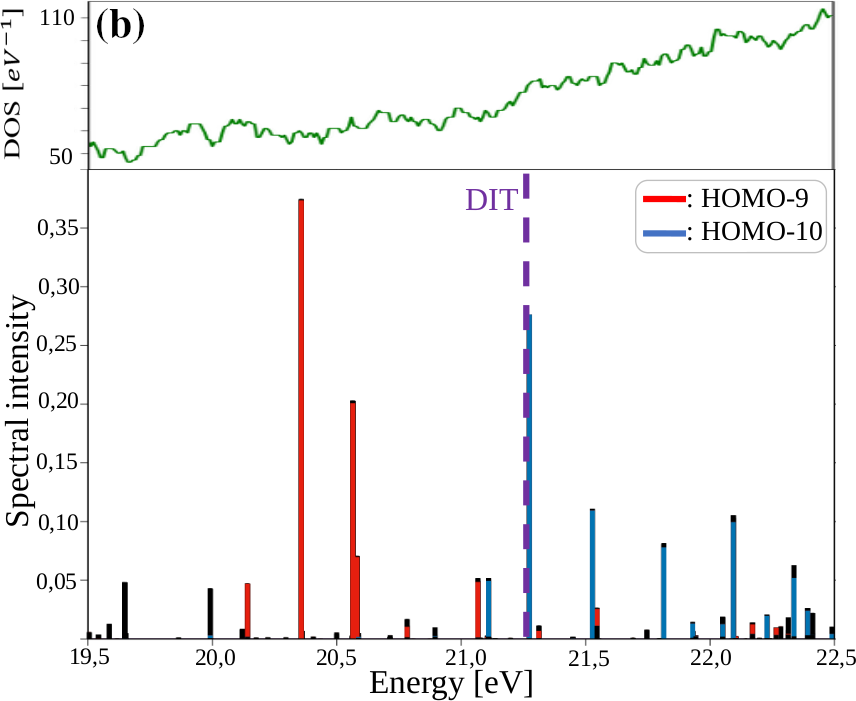}
    \label{fig:spec_bc}
  \end{subfigure}
\caption{\label{fig:o1}\textbf{(a)} Density of states (top panel) and nD-ADC(3) ionization spectrum (lower panel) of 1,3,5,7,9-Decapentayne (C$_{10}$H$_2$). Contributions of the 1h configurations corresponding to an electron missing in HOMO$-9$ and in HOMO$-10$ are depicted in red and blue, respectively. The energy of the lowest double-ionization potential (DIT) computed with pp-ADC(2) is denoted by a dashed purple vertical line. 
\textbf{(b)} Zoom of the correlation-band region of C$_{10}$H$_2$. DOS stands for Density Of States.}
\end{figure}

To illustrate our results we will take 1,3,5,7,9-Decapentayne (C$_{10}$H$_2$) as an example. In Fig.~\ref{fig:o1} the nD-ADC(3) ionization spectrum of C$_{10}$H$_2$ is depicted. In this spectrum each vertical line corresponds to a cationic eigenstate. The position of the line on the abscissa gives the corresponding ionization energy and its height the spectral intensity of the state (which correspond to the weight of all 1h contributions). What is missing to reach a value of 1 corresponds to the contribution of the 2h1p configurations. In the lowest energy range, most of the states have strong 1h character, whereas the more we go higher in energy and approach the Double Ionization Threshold (DIT), the more the 2h1p character starts to dominate in the spectrum. In fact, close to the DIT at 21~eV (calculated with the two-particle ADC(2) \cite{schirmer1984higher} and shown with purple dashed line in Fig.~\ref{fig:o1}) there is a quasi-continuum of states where the 1h contributions almost disappear. It is no longer possible to distinguish which line is the main state and which lines are its satellites, and to assign the states using only a few electronic configurations. This phenomenon is called the breakdown of the molecular-orbital picture of ionization \cite{cederbaum1999ultrafast} and leads to the creation of the CB.

We would like first to discuss the shape of the DOS in the CB. In Ref.~\cite{herve2021ultrafast}, it was shown that for PAH molecules the DOS evolves linearly as a function of energy in the CB region, but how general is this behavior? In Fig~\ref{fig:o1}(b) we see that in the energy range corresponding to the CB of C$_{10}$H$_2$ the evolution of DOS is indeed linear. This is observed for all the molecules studied in this paper, with a behavior becoming closer and closer to linear when the size of the systems increases. Moreover, as we see in Fig.~\ref{fig:o1}(a), after the DIT the DOS increases at a faster rate, which is due to the ionization continuum, which gets discretized within the nD-ADC(3) method. As can be seen in Fig.~\ref{fig:o1}(b), the CB is created mainly from states resulting from the ionization of two molecular orbitals, HOMO$-9$ and HOMO$-10$, depicted in red and blue, respectively. States with 1h contribution mainly from HOMO$-9$ are mostly bellow the DIT, while the states with mainly HOMO$-10$ contribution are mostly above the DIT. This is a general behavior that is observed in all the molecules studied in this work. In the following, we will refer to these two orbitals as the ``above-DIT'' and ``below-DIT'' orbitals when properties applying to all molecules are discussed.

Now, let us focus on the correlation-driven charge migrations triggered by the removal of an electron from one of those two orbitals. To do so, we used the sudden-ionization approximation, assuming that at $t=0$ an electron is removed from a given orbital. The created initial state was then propagated using the multielectron wave-packet propagation method \cite{kuleff2005multielectron}. To characterize the triggered charge-migration dynamics, we followed in time and space the density of the hole created in the electronic cloud by the ionization. The latter can be defined as \cite{cederbaum1999ultrafast,breidbach2003migration}:
\begin{align}
    \mathcal{Q}(\vec{r},t) &=\bra{\Psi_0}\hat{\rho}(\vec{r})\ket{\Psi_0}-\bra{\Phi_i(t)}(\hat{\rho}(\vec{r})\ket{\Phi_i(t)}\nonumber \\
    &=\rho_0(\vec{r})-\rho_i(\vec{r},t),
\end{align}
where $\ket{\Psi_0}$ is the ground state of the neutral molecule and $\ket{\Phi_i(t)}$ is the multielectron wave packet of the created ion, while $\hat{\rho}$ is the position-dependent electronic density operator. The hole density is thus the difference between the electron density in the neutral ground state $\rho_0(\vec{r})$ and the time-dependent cationic density $\rho_i(\vec{r},t)$. For further details, see Refs.~\cite{breidbach2003migration,kuleff2018ultrafast}. 

\begin{figure}[!ht]
  \begin{subfigure}{}
    \centering
    \includegraphics[scale=0.45]{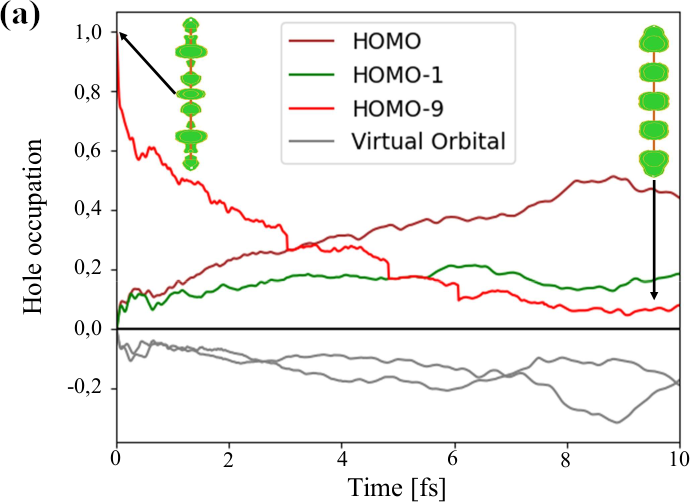} 
    \label{fig:homo10}
  \end{subfigure}
  \begin{subfigure}{}
    \centering
    \includegraphics[scale=0.45]{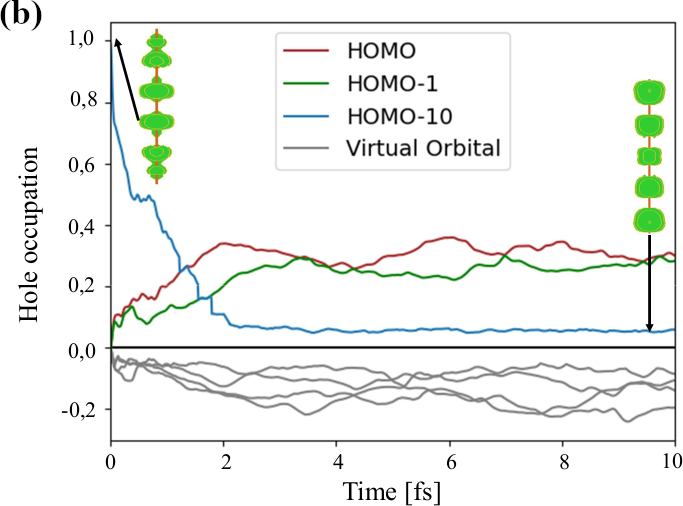}
    \label{fig:homo11}
  \end{subfigure}
\caption{\label{fig:o2} 
\textbf{(a)}: Hole occupation for the ionization of HOMO$-9$.
\textbf{(b)}: Hole occupation for the ionization of HOMO$-10$.
Negative hole occupation means that an electron is promoted to a virtual orbital. The hole densities at 0 and 10~fs are also depicted.
} 
\end{figure}

The evolution of the hole density can be analyzed in terms of the time-dependent hole-occupation numbers, describing the part of the hole being in a given orbital at time $t$ \cite{cederbaum1999ultrafast,breidbach2003migration,kuleff2005multielectron}. The evolution of the hole occupations after the ionization of HOMO$-9$ and HOMO$-10$ are presented in Fig.~\ref{fig:o2}. By removing and electron from one of these orbitals, all states containing a 1h contribution from the corresponding orbital are populated. The amount of population is given by the weight of the 1h configuration in the state. It is important to note here that within the present methodology the nuclei are frozen and, therefore, no transfer of population between the states can take place due to non-adiabatic effects. The dynamics thus results solely from phase difference between the populated states, or, in other words, from the created coherence. 

As we can see, at $t=0$ the population of the initially ionized orbital is equal to 1. As time proceeds, this occupation number decreases quickly, while new holes are open in other occupied orbitals and electrons are created in virtual orbitals (negative hole occupation corresponds to population of a virtual orbital). Such populations are made possible due to the many 2h1p configurations contributing to the initially populated cationic states. The decrease of the population of the initial orbital has an exponential shape. This behavior was predicted and analyzed in Ref.~\cite{breidbach2003migration}. Using a least-square-fit procedure with an exponential function, we extracted the relaxation time for the above and below-DIT holes of all studied molecules. We observed that the relaxation time is very different for the ionization of HOMO$-9$ and HOMO$-10$ for C$_{10}$H$_2$ and HOMO$-10$ and HOMO$-11$ of C$_{12}$H$_2$. The hole in below-DIT orbital has a much longer relaxation time than that in the above one. This is a consequence of the increase of the DOS above the DIT due to the ionization continuum and the possibility for the systems to undergo an Auger decay. For smaller systems in the studied series of molecules above- and below-DIT orbitals have comparable relaxation time (see, Tab.~\ref{tab:tau}). We note that the multielectron wave-packet propagation technique can also describe electronic decay processes and was successfully used for that in the past \cite{Kuleff2007Tracing,kuleff2017electronic,Mullenix2020Electronic}.

\begin{table}
\centering
\caption{Relaxation time for above- and below-DIT orbitals in the correlation band for each system.}
\label{tab:tau}
\begin{tabular}{cccc}
\hline \hline 
Molecule & Orbital & Relaxation time [fs]  \\
\hline \hline
\multirow{2}{*}{$C_4H_2$} & HOMO-6 & 0.34 \\
    & HOMO-5 & 0.28 &  \\ \hline
\multirow{2}{*}{$C_6H_2$} & HOMO-8 & 0.25 \\
    & HOMO-7 & 0.81 & \\ \hline
\multirow{2}{*}{$C_8H_2$} & HOMO-9 & 0.80  \\
    & HOMO-8 & 1.15 & \\   \hline
\multirow{2}{*}{$C_{10}H_2$} & HOMO-10 & 0.97 \\
    & HOMO-9 & 3.76 & \\ \hline
\multirow{2}{*}{$C_{12}H_2$} & HOMO-11 & 2.30 \\
    & HOMO-10 & 7.10  \\ 
\hline\hline
\end{tabular}
\end{table}

\begin{figure}[!ht]
  \begin{subfigure}{}
    \centering
    \includegraphics[scale=0.54]{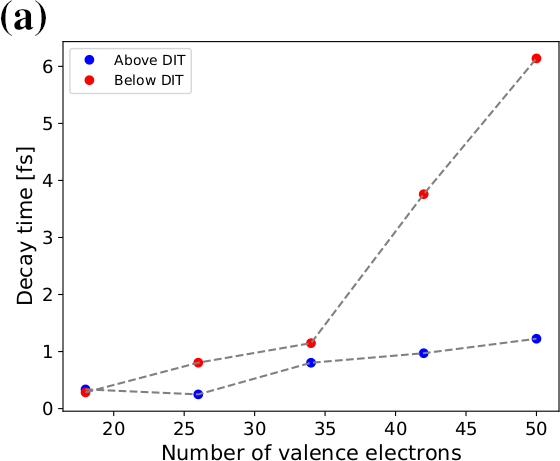}
    \label{fig:tau}
  \end{subfigure}
  \begin{subfigure}
    \centering
    \includegraphics[scale=0.54]{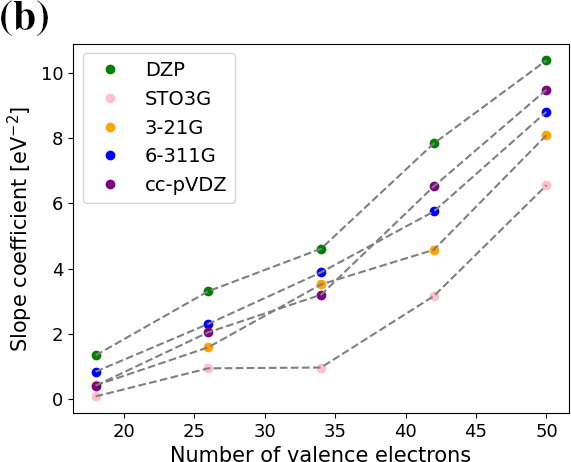}
    \label{fig:desnite}
  \end{subfigure}
\caption{\label{fig:o3} \textbf{(a)}: Decay time for each alkyne. In blue: the decay for the orbital above the DIT. In red, the one below the DIT
\textbf{(b)}: Slope gradient of the linear fitting of the density of states in the correlation band for each alkyne }
\end{figure}

Let us now discuss the possible scaling law for both the time scale of the correlation-driven charge migration and the slope of the DOS in the CB. In Fig.~\ref{fig:o3}(a) we present the relaxation time for each alkyne chain for both, below- and above-DIT orbitals that constitute the CB. The green curve represents the evolution of the relaxation time for the below-DIT orbitals and the orange is for the above-DIT orbitals. We notice that for each system, except the smallest one, the relaxation time is longer for the below-DIT orbitals as previously mentioned. Moreover, we clearly see that for both types of orbitals, the relaxation time increases with the size of the alkyne chains. However, we clearly see that the green and the orange curves have a different behavior. The above-DIT orbital relaxation slowly increases with the size of the systems in nearly linear manner. The below-DIT relaxation behaves similarly than the above-DIT one for the small system and then significantly increases for the larger ones. The time scales obtained for larger systems, therefore, appear more suitable for control and thus more promising for both the development of a molecular design based on electron correlation and attochemistry. This difference of behavior between above- and bellow-DIT orbitals can be explained by the physical processes that are initiated. As we saw, the ionization from the above-DIT orbitals populates also states that can relax through an Auger decay and thus some part of the observed relaxation time reflects actually the electronic decay time of these states. It is, however, not easy to extract and quantify the electronic decay component, as both the cationic states and the DIT are computed with some error which makes difficult the separation of the populated states in decaying and non-decaying ones.

What can we say about the increase of the DOS with energy? We discussed already that the DOS in the CB increases linearly in all studied molecules. As the states in the CB have predominantly 2h1p character, it is clear that the number of states that we obtain is sensitive to the basis sets used. A larger basis provides a larger number of virtual orbitals and thus in general a better description of the correlation effects. Is the DOS and respectively the slope of its increase obtained with not very large basis sets of some use?

In Fig.~\ref{fig:o3}(b) we report the slope coefficients (from a linear fitting) for the DOS in the CB for each system obtained from calculations with several different basis set. We see that although the slope for each of the systems sensitively depends on the basis set used, the overall trend as a function of the system size is correctly reproduced by all basis sets (probably apart from the smallest STO3G). We see that the slope increases with the size of the system and that this increase is nearly linear as function of the number of valence electrons. It is important to note that even small basis sets as 3-21G correctly reproduce the overall behavior of the evolution of the DOS slope as a function of the number of valence electrons. As this slope is one of the 3 main parameters of the model of Ref.~\cite{herve2021ultrafast}, it would be of interest to be able to predict the slope of the DOS of system of large size without performing expensive calculations. The present results open the way to obtain such a value by extrapolating the results for smaller systems obtained even with small basis sets. Moreover, we can largely improve the value by saturating the number of states for the smallest system of the molecular family by increasing the size of the basis. In such a way, we can get a correction factor for the DOS slope in large systems of the family for which calculations are prohibitively expensive.      

\section{Conclusion and outlook}

In this paper, we studied the scaling laws characterizing the relaxation in the correlation bands of alkynes chains. We observed that the relaxation time of the correlation-driven charge migration initiated by ionization of orbitals populating states below the double ionization threshold increases with the size of the systems, reaching several femtoseconds for the two largest systems studied. These are time scales that can be experimentally accessible. As the dynamics in the correlation band have been shown to impact the stability of molecular systems \cite{maansson2021real,despre2022correlation}, the next step in the context of attochemistry would be to search for control schemes for charge migration \cite{golubev2015control} adapted for high density of states.

We also observed that the slope of the density of states in the correlation band increases nearly linearly with the number of valence electrons, which allows one to estimate this parameter for larger systems in the molecular family. This is a key parameter of the model able to predict the non-adiabatic relaxation time in the CB of a given molecular family \cite{herve2021ultrafast}. The model, however, depends also on the average non-adiabatic coupling in the CB and the averaged frequency of the vibrational mode that couples the states in the CB. It would be, therefore, interesting to investigate the behavior of these parameters as a function of system size and search for similar scaling laws. This would necessitate to perform non-adiabatic dynamics simulations for several systems from the same molecular family. We hope that our work will stimulate such endeavor.

\begin{acknowledgments}
AIK and VD acknowledges financial support from the ANR-DFG project FAUST.
\end{acknowledgments}

\bibliography{alkyne_CB.bib}

\begin{thebibliography}{58}%
\makeatletter
\providecommand \@ifxundefined [1]{%
 \@ifx{#1\undefined}
}%
\providecommand \@ifnum [1]{%
 \ifnum #1\expandafter \@firstoftwo
 \else \expandafter \@secondoftwo
 \fi
}%
\providecommand \@ifx [1]{%
 \ifx #1\expandafter \@firstoftwo
 \else \expandafter \@secondoftwo
 \fi
}%
\providecommand \natexlab [1]{#1}%
\providecommand \enquote  [1]{``#1''}%
\providecommand \bibnamefont  [1]{#1}%
\providecommand \bibfnamefont [1]{#1}%
\providecommand \citenamefont [1]{#1}%
\providecommand \href@noop [0]{\@secondoftwo}%
\providecommand \href [0]{\begingroup \@sanitize@url \@href}%
\providecommand \@href[1]{\@@startlink{#1}\@@href}%
\providecommand \@@href[1]{\endgroup#1\@@endlink}%
\providecommand \@sanitize@url [0]{\catcode `\\12\catcode `\$12\catcode `\&12\catcode `\#12\catcode `\^12\catcode `\_12\catcode `\%12\relax}%
\providecommand \@@startlink[1]{}%
\providecommand \@@endlink[0]{}%
\providecommand \url  [0]{\begingroup\@sanitize@url \@url }%
\providecommand \@url [1]{\endgroup\@href {#1}{\urlprefix }}%
\providecommand \urlprefix  [0]{URL }%
\providecommand \Eprint [0]{\href }%
\providecommand \doibase [0]{http://dx.doi.org/}%
\providecommand \selectlanguage [0]{\@gobble}%
\providecommand \bibinfo  [0]{\@secondoftwo}%
\providecommand \bibfield  [0]{\@secondoftwo}%
\providecommand \translation [1]{[#1]}%
\providecommand \BibitemOpen [0]{}%
\providecommand \bibitemStop [0]{}%
\providecommand \bibitemNoStop [0]{.\EOS\space}%
\providecommand \EOS [0]{\spacefactor3000\relax}%
\providecommand \BibitemShut  [1]{\csname bibitem#1\endcsname}%
\let\auto@bib@innerbib\@empty
\bibitem [{\citenamefont {Nisoli}\ \emph {et~al.}(2017)\citenamefont {Nisoli}, \citenamefont {Decleva}, \citenamefont {Calegari}, \citenamefont {Palacios},\ and\ \citenamefont {Mart{\'\i}n}}]{nisoli2017attosecond}%
  \BibitemOpen
  \bibfield  {author} {\bibinfo {author} {\bibfnamefont {M.}~\bibnamefont {Nisoli}}, \bibinfo {author} {\bibfnamefont {P.}~\bibnamefont {Decleva}}, \bibinfo {author} {\bibfnamefont {F.}~\bibnamefont {Calegari}}, \bibinfo {author} {\bibfnamefont {A.}~\bibnamefont {Palacios}}, \ and\ \bibinfo {author} {\bibfnamefont {F.}~\bibnamefont {Mart{\'\i}n}},\ }\href@noop {} {\bibfield  {journal} {\bibinfo  {journal} {Chem. Rev.}\ }\textbf {\bibinfo {volume} {117}},\ \bibinfo {pages} {10760} (\bibinfo {year} {2017})}\BibitemShut {NoStop}%
\bibitem [{\citenamefont {L{\'e}pine}\ \emph {et~al.}(2014)\citenamefont {L{\'e}pine}, \citenamefont {Ivanov},\ and\ \citenamefont {Vrakking}}]{lepine2014attosecond}%
  \BibitemOpen
  \bibfield  {author} {\bibinfo {author} {\bibfnamefont {F.}~\bibnamefont {L{\'e}pine}}, \bibinfo {author} {\bibfnamefont {M.~Y.}\ \bibnamefont {Ivanov}}, \ and\ \bibinfo {author} {\bibfnamefont {M.~J.}\ \bibnamefont {Vrakking}},\ }\href@noop {} {\bibfield  {journal} {\bibinfo  {journal} {Nat. Photonics}\ }\textbf {\bibinfo {volume} {8}},\ \bibinfo {pages} {195} (\bibinfo {year} {2014})}\BibitemShut {NoStop}%
\bibitem [{\citenamefont {Kling}\ \emph {et~al.}(2006)\citenamefont {Kling}, \citenamefont {Siedschlag}, \citenamefont {Verhoef}, \citenamefont {Khan}, \citenamefont {Schultze}, \citenamefont {Uphues}, \citenamefont {Ni}, \citenamefont {Uiberacker}, \citenamefont {Drescher}, \citenamefont {Krausz} \emph {et~al.}}]{kling2006control}%
  \BibitemOpen
  \bibfield  {author} {\bibinfo {author} {\bibfnamefont {M.}~\bibnamefont {Kling}}, \bibinfo {author} {\bibfnamefont {C.}~\bibnamefont {Siedschlag}}, \bibinfo {author} {\bibfnamefont {A.~J.}\ \bibnamefont {Verhoef}}, \bibinfo {author} {\bibfnamefont {J.}~\bibnamefont {Khan}}, \bibinfo {author} {\bibfnamefont {M.}~\bibnamefont {Schultze}}, \bibinfo {author} {\bibfnamefont {T.}~\bibnamefont {Uphues}}, \bibinfo {author} {\bibfnamefont {Y.}~\bibnamefont {Ni}}, \bibinfo {author} {\bibfnamefont {M.}~\bibnamefont {Uiberacker}}, \bibinfo {author} {\bibfnamefont {M.}~\bibnamefont {Drescher}}, \bibinfo {author} {\bibfnamefont {F.}~\bibnamefont {Krausz}},  \emph {et~al.},\ }\href@noop {} {\bibfield  {journal} {\bibinfo  {journal} {Science}\ }\textbf {\bibinfo {volume} {312}},\ \bibinfo {pages} {246} (\bibinfo {year} {2006})}\BibitemShut {NoStop}%
\bibitem [{\citenamefont {Neidel}\ \emph {et~al.}(2013)\citenamefont {Neidel}, \citenamefont {Klei}, \citenamefont {Yang}, \citenamefont {Rouz{\'e}e}, \citenamefont {Vrakking}, \citenamefont {Kl{\"u}nder}, \citenamefont {Miranda}, \citenamefont {Arnold}, \citenamefont {Fordell}, \citenamefont {L’Huillier} \emph {et~al.}}]{neidel2013probing}%
  \BibitemOpen
  \bibfield  {author} {\bibinfo {author} {\bibfnamefont {C.}~\bibnamefont {Neidel}}, \bibinfo {author} {\bibfnamefont {J.}~\bibnamefont {Klei}}, \bibinfo {author} {\bibfnamefont {C.-H.}\ \bibnamefont {Yang}}, \bibinfo {author} {\bibfnamefont {A.}~\bibnamefont {Rouz{\'e}e}}, \bibinfo {author} {\bibfnamefont {M.}~\bibnamefont {Vrakking}}, \bibinfo {author} {\bibfnamefont {K.}~\bibnamefont {Kl{\"u}nder}}, \bibinfo {author} {\bibfnamefont {M.}~\bibnamefont {Miranda}}, \bibinfo {author} {\bibfnamefont {C.}~\bibnamefont {Arnold}}, \bibinfo {author} {\bibfnamefont {T.}~\bibnamefont {Fordell}}, \bibinfo {author} {\bibfnamefont {A.}~\bibnamefont {L’Huillier}},  \emph {et~al.},\ }\href@noop {} {\bibfield  {journal} {\bibinfo  {journal} {Phys. Rev. Lett.}\ }\textbf {\bibinfo {volume} {111}},\ \bibinfo {pages} {033001} (\bibinfo {year} {2013})}\BibitemShut {NoStop}%
\bibitem [{\citenamefont {Haessler}\ \emph {et~al.}(2009)\citenamefont {Haessler}, \citenamefont {Fabre}, \citenamefont {Higuet}, \citenamefont {Caillat}, \citenamefont {Ruchon}, \citenamefont {Breger}, \citenamefont {Carr{\'e}}, \citenamefont {Constant}, \citenamefont {Maquet}, \citenamefont {M{\'e}vel} \emph {et~al.}}]{haessler2009phase}%
  \BibitemOpen
  \bibfield  {author} {\bibinfo {author} {\bibfnamefont {S.}~\bibnamefont {Haessler}}, \bibinfo {author} {\bibfnamefont {B.}~\bibnamefont {Fabre}}, \bibinfo {author} {\bibfnamefont {J.}~\bibnamefont {Higuet}}, \bibinfo {author} {\bibfnamefont {J.}~\bibnamefont {Caillat}}, \bibinfo {author} {\bibfnamefont {T.}~\bibnamefont {Ruchon}}, \bibinfo {author} {\bibfnamefont {P.}~\bibnamefont {Breger}}, \bibinfo {author} {\bibfnamefont {B.}~\bibnamefont {Carr{\'e}}}, \bibinfo {author} {\bibfnamefont {E.}~\bibnamefont {Constant}}, \bibinfo {author} {\bibfnamefont {A.}~\bibnamefont {Maquet}}, \bibinfo {author} {\bibfnamefont {E.}~\bibnamefont {M{\'e}vel}},  \emph {et~al.},\ }\href@noop {} {\bibfield  {journal} {\bibinfo  {journal} {Phys. Rev. A}\ }\textbf {\bibinfo {volume} {80}},\ \bibinfo {pages} {011404} (\bibinfo {year} {2009})}\BibitemShut {NoStop}%
\bibitem [{\citenamefont {Calegari}\ \emph {et~al.}(2014)\citenamefont {Calegari}, \citenamefont {Ayuso}, \citenamefont {Trabattoni}, \citenamefont {Belshaw}, \citenamefont {De~Camillis}, \citenamefont {Anumula}, \citenamefont {Frassetto}, \citenamefont {Poletto}, \citenamefont {Palacios}, \citenamefont {Decleva} \emph {et~al.}}]{calegari2014ultrafast}%
  \BibitemOpen
  \bibfield  {author} {\bibinfo {author} {\bibfnamefont {F.}~\bibnamefont {Calegari}}, \bibinfo {author} {\bibfnamefont {D.}~\bibnamefont {Ayuso}}, \bibinfo {author} {\bibfnamefont {A.}~\bibnamefont {Trabattoni}}, \bibinfo {author} {\bibfnamefont {L.}~\bibnamefont {Belshaw}}, \bibinfo {author} {\bibfnamefont {S.}~\bibnamefont {De~Camillis}}, \bibinfo {author} {\bibfnamefont {S.}~\bibnamefont {Anumula}}, \bibinfo {author} {\bibfnamefont {F.}~\bibnamefont {Frassetto}}, \bibinfo {author} {\bibfnamefont {L.}~\bibnamefont {Poletto}}, \bibinfo {author} {\bibfnamefont {A.}~\bibnamefont {Palacios}}, \bibinfo {author} {\bibfnamefont {P.}~\bibnamefont {Decleva}},  \emph {et~al.},\ }\href@noop {} {\bibfield  {journal} {\bibinfo  {journal} {Science}\ }\textbf {\bibinfo {volume} {346}},\ \bibinfo {pages} {336} (\bibinfo {year} {2014})}\BibitemShut {NoStop}%
\bibitem [{\citenamefont {Marciniak}\ \emph {et~al.}(2019)\citenamefont {Marciniak}, \citenamefont {Despr{\'e}}, \citenamefont {Loriot}, \citenamefont {Karras}, \citenamefont {Herv{\'e}}, \citenamefont {Quintard}, \citenamefont {Catoire}, \citenamefont {Joblin}, \citenamefont {Constant}, \citenamefont {Kuleff} \emph {et~al.}}]{marciniak2019electron}%
  \BibitemOpen
  \bibfield  {author} {\bibinfo {author} {\bibfnamefont {A.}~\bibnamefont {Marciniak}}, \bibinfo {author} {\bibfnamefont {V.}~\bibnamefont {Despr{\'e}}}, \bibinfo {author} {\bibfnamefont {V.}~\bibnamefont {Loriot}}, \bibinfo {author} {\bibfnamefont {G.}~\bibnamefont {Karras}}, \bibinfo {author} {\bibfnamefont {M.}~\bibnamefont {Herv{\'e}}}, \bibinfo {author} {\bibfnamefont {L.}~\bibnamefont {Quintard}}, \bibinfo {author} {\bibfnamefont {F.}~\bibnamefont {Catoire}}, \bibinfo {author} {\bibfnamefont {C.}~\bibnamefont {Joblin}}, \bibinfo {author} {\bibfnamefont {E.}~\bibnamefont {Constant}}, \bibinfo {author} {\bibfnamefont {A.}~\bibnamefont {Kuleff}},  \emph {et~al.},\ }\href@noop {} {\bibfield  {journal} {\bibinfo  {journal} {Nat. Commun.}\ }\textbf {\bibinfo {volume} {10}},\ \bibinfo {pages} {337} (\bibinfo {year} {2019})}\BibitemShut {NoStop}%
\bibitem [{\citenamefont {Kuleff}\ \emph {et~al.}(2010)\citenamefont {Kuleff}, \citenamefont {L\"unnemann},\ and\ \citenamefont {Cederbaum}}]{kuleff2010ultrafast}%
  \BibitemOpen
  \bibfield  {author} {\bibinfo {author} {\bibfnamefont {A.~I.}\ \bibnamefont {Kuleff}}, \bibinfo {author} {\bibfnamefont {S.}~\bibnamefont {L\"unnemann}}, \ and\ \bibinfo {author} {\bibfnamefont {L.~S.}\ \bibnamefont {Cederbaum}},\ }\href@noop {} {\bibfield  {journal} {\bibinfo  {journal} {J. Phys. Chem. A}\ }\textbf {\bibinfo {volume} {114}},\ \bibinfo {pages} {8676} (\bibinfo {year} {2010})}\BibitemShut {NoStop}%
\bibitem [{\citenamefont {Kuleff}\ \emph {et~al.}(2016)\citenamefont {Kuleff}, \citenamefont {Kryzhevoi}, \citenamefont {Pernpointner},\ and\ \citenamefont {Cederbaum}}]{kuleff2016core}%
  \BibitemOpen
  \bibfield  {author} {\bibinfo {author} {\bibfnamefont {A.~I.}\ \bibnamefont {Kuleff}}, \bibinfo {author} {\bibfnamefont {N.~V.}\ \bibnamefont {Kryzhevoi}}, \bibinfo {author} {\bibfnamefont {M.}~\bibnamefont {Pernpointner}}, \ and\ \bibinfo {author} {\bibfnamefont {L.~S.}\ \bibnamefont {Cederbaum}},\ }\href@noop {} {\bibfield  {journal} {\bibinfo  {journal} {Phys. Rev. Lett.}\ }\textbf {\bibinfo {volume} {117}},\ \bibinfo {pages} {093002} (\bibinfo {year} {2016})}\BibitemShut {NoStop}%
\bibitem [{\citenamefont {Mignolet}\ \emph {et~al.}(2014)\citenamefont {Mignolet}, \citenamefont {Levine},\ and\ \citenamefont {Remacle}}]{mignolet2014charge}%
  \BibitemOpen
  \bibfield  {author} {\bibinfo {author} {\bibfnamefont {B.}~\bibnamefont {Mignolet}}, \bibinfo {author} {\bibfnamefont {R.~D.}\ \bibnamefont {Levine}}, \ and\ \bibinfo {author} {\bibfnamefont {F.}~\bibnamefont {Remacle}},\ }\href@noop {} {\bibfield  {journal} {\bibinfo  {journal} {J. Phys. B: At. Mol. Opt. Phys.}\ }\textbf {\bibinfo {volume} {47}},\ \bibinfo {pages} {124011} (\bibinfo {year} {2014})}\BibitemShut {NoStop}%
\bibitem [{\citenamefont {Sun}\ \emph {et~al.}(2017)\citenamefont {Sun}, \citenamefont {Mignolet}, \citenamefont {Fan}, \citenamefont {Li}, \citenamefont {Levine},\ and\ \citenamefont {Remacle}}]{sun2017nuclear}%
  \BibitemOpen
  \bibfield  {author} {\bibinfo {author} {\bibfnamefont {S.}~\bibnamefont {Sun}}, \bibinfo {author} {\bibfnamefont {B.}~\bibnamefont {Mignolet}}, \bibinfo {author} {\bibfnamefont {L.}~\bibnamefont {Fan}}, \bibinfo {author} {\bibfnamefont {W.}~\bibnamefont {Li}}, \bibinfo {author} {\bibfnamefont {R.~D.}\ \bibnamefont {Levine}}, \ and\ \bibinfo {author} {\bibfnamefont {F.}~\bibnamefont {Remacle}},\ }\href@noop {} {\bibfield  {journal} {\bibinfo  {journal} {J. Phys. Chem. A}\ }\textbf {\bibinfo {volume} {121}},\ \bibinfo {pages} {1442} (\bibinfo {year} {2017})}\BibitemShut {NoStop}%
\bibitem [{\citenamefont {Lara-Astiaso}\ \emph {et~al.}(2016)\citenamefont {Lara-Astiaso}, \citenamefont {Ayuso}, \citenamefont {Tavernelli}, \citenamefont {Decleva}, \citenamefont {Palacios},\ and\ \citenamefont {Mart\'{\i}n}}]{lara-astiaso2016decoherence}%
  \BibitemOpen
  \bibfield  {author} {\bibinfo {author} {\bibfnamefont {M.}~\bibnamefont {Lara-Astiaso}}, \bibinfo {author} {\bibfnamefont {D.}~\bibnamefont {Ayuso}}, \bibinfo {author} {\bibfnamefont {I.}~\bibnamefont {Tavernelli}}, \bibinfo {author} {\bibfnamefont {P.}~\bibnamefont {Decleva}}, \bibinfo {author} {\bibfnamefont {A.}~\bibnamefont {Palacios}}, \ and\ \bibinfo {author} {\bibfnamefont {F.}~\bibnamefont {Mart\'{\i}n}},\ }\href@noop {} {\bibfield  {journal} {\bibinfo  {journal} {Faraday Discuss.}\ }\textbf {\bibinfo {volume} {194}},\ \bibinfo {pages} {41} (\bibinfo {year} {2016})}\BibitemShut {NoStop}%
\bibitem [{\citenamefont {Lara-Astiaso}\ \emph {et~al.}(2018{\natexlab{a}})\citenamefont {Lara-Astiaso}, \citenamefont {Galli}, \citenamefont {Trabattoni}, \citenamefont {Palacios}, \citenamefont {Ayuso}, \citenamefont {Frassetto}, \citenamefont {Poletto}, \citenamefont {De~Camillis}, \citenamefont {Greenwood}, \citenamefont {Decleva}, \citenamefont {Tavernelli}, \citenamefont {Calegari}, \citenamefont {Nisoli},\ and\ \citenamefont {Mart\'{\i}n}}]{lara-astiaso2018attosecond}%
  \BibitemOpen
  \bibfield  {author} {\bibinfo {author} {\bibfnamefont {M.}~\bibnamefont {Lara-Astiaso}}, \bibinfo {author} {\bibfnamefont {M.}~\bibnamefont {Galli}}, \bibinfo {author} {\bibfnamefont {A.}~\bibnamefont {Trabattoni}}, \bibinfo {author} {\bibfnamefont {A.}~\bibnamefont {Palacios}}, \bibinfo {author} {\bibfnamefont {D.}~\bibnamefont {Ayuso}}, \bibinfo {author} {\bibfnamefont {F.}~\bibnamefont {Frassetto}}, \bibinfo {author} {\bibfnamefont {L.}~\bibnamefont {Poletto}}, \bibinfo {author} {\bibfnamefont {S.}~\bibnamefont {De~Camillis}}, \bibinfo {author} {\bibfnamefont {J.}~\bibnamefont {Greenwood}}, \bibinfo {author} {\bibfnamefont {P.}~\bibnamefont {Decleva}}, \bibinfo {author} {\bibfnamefont {I.}~\bibnamefont {Tavernelli}}, \bibinfo {author} {\bibfnamefont {F.}~\bibnamefont {Calegari}}, \bibinfo {author} {\bibfnamefont {M.}~\bibnamefont {Nisoli}}, \ and\ \bibinfo {author} {\bibfnamefont {F.}~\bibnamefont {Mart\'{\i}n}},\ }\href@noop {} {\bibfield  {journal} {\bibinfo  {journal} {J. Phys. Chem. Lett.}\ }\textbf
  {\bibinfo {volume} {9}},\ \bibinfo {pages} {4570} (\bibinfo {year} {2018}{\natexlab{a}})}\BibitemShut {NoStop}%
\bibitem [{\citenamefont {Yuan}\ and\ \citenamefont {Bandrauk}(2019)}]{yuan2019ultrafast}%
  \BibitemOpen
  \bibfield  {author} {\bibinfo {author} {\bibfnamefont {K.}~\bibnamefont {Yuan}}\ and\ \bibinfo {author} {\bibfnamefont {A.~D.}\ \bibnamefont {Bandrauk}},\ }\href@noop {} {\bibfield  {journal} {\bibinfo  {journal} {J. Phys. Chem. A}\ } (\bibinfo {year} {2019})}\BibitemShut {NoStop}%
\bibitem [{\citenamefont {Folorunso}\ \emph {et~al.}(2023)\citenamefont {Folorunso}, \citenamefont {Mauger}, \citenamefont {Hamer}, \citenamefont {Jayasinghe}, \citenamefont {Wahyutama}, \citenamefont {Ragains}, \citenamefont {Jones}, \citenamefont {DiMauro}, \citenamefont {Gaarde}, \citenamefont {Schafer} \emph {et~al.}}]{folorunso2023attochemistry}%
  \BibitemOpen
  \bibfield  {author} {\bibinfo {author} {\bibfnamefont {A.~S.}\ \bibnamefont {Folorunso}}, \bibinfo {author} {\bibfnamefont {F.}~\bibnamefont {Mauger}}, \bibinfo {author} {\bibfnamefont {K.~A.}\ \bibnamefont {Hamer}}, \bibinfo {author} {\bibfnamefont {D.~D.}\ \bibnamefont {Jayasinghe}}, \bibinfo {author} {\bibfnamefont {I.~S.}\ \bibnamefont {Wahyutama}}, \bibinfo {author} {\bibfnamefont {J.~R.}\ \bibnamefont {Ragains}}, \bibinfo {author} {\bibfnamefont {R.~R.}\ \bibnamefont {Jones}}, \bibinfo {author} {\bibfnamefont {L.~F.}\ \bibnamefont {DiMauro}}, \bibinfo {author} {\bibfnamefont {M.~B.}\ \bibnamefont {Gaarde}}, \bibinfo {author} {\bibfnamefont {K.~J.}\ \bibnamefont {Schafer}},  \emph {et~al.},\ }\href@noop {} {\bibfield  {journal} {\bibinfo  {journal} {J. Phys. Chem. A}\ }\textbf {\bibinfo {volume} {127}},\ \bibinfo {pages} {1894} (\bibinfo {year} {2023})}\BibitemShut {NoStop}%
\bibitem [{\citenamefont {Haase}\ \emph {et~al.}(2021)\citenamefont {Haase}, \citenamefont {Hermann}, \citenamefont {Manz}, \citenamefont {Pohl},\ and\ \citenamefont {Tremblay}}]{haase2021electron}%
  \BibitemOpen
  \bibfield  {author} {\bibinfo {author} {\bibfnamefont {D.}~\bibnamefont {Haase}}, \bibinfo {author} {\bibfnamefont {G.}~\bibnamefont {Hermann}}, \bibinfo {author} {\bibfnamefont {J.}~\bibnamefont {Manz}}, \bibinfo {author} {\bibfnamefont {V.}~\bibnamefont {Pohl}}, \ and\ \bibinfo {author} {\bibfnamefont {J.~C.}\ \bibnamefont {Tremblay}},\ }\href@noop {} {\bibfield  {journal} {\bibinfo  {journal} {Symmetry}\ }\textbf {\bibinfo {volume} {13}},\ \bibinfo {pages} {205} (\bibinfo {year} {2021})}\BibitemShut {NoStop}%
\bibitem [{\citenamefont {Cederbaum}\ and\ \citenamefont {Zobeley}(1999)}]{cederbaum1999ultrafast}%
  \BibitemOpen
  \bibfield  {author} {\bibinfo {author} {\bibfnamefont {L.~S.}\ \bibnamefont {Cederbaum}}\ and\ \bibinfo {author} {\bibfnamefont {J.}~\bibnamefont {Zobeley}},\ }\href@noop {} {\bibfield  {journal} {\bibinfo  {journal} {Chem. Phys. Lett.}\ }\textbf {\bibinfo {volume} {307}},\ \bibinfo {pages} {205} (\bibinfo {year} {1999})}\BibitemShut {NoStop}%
\bibitem [{\citenamefont {Kuleff}\ and\ \citenamefont {Cederbaum}(2014)}]{kuleff2014ultrafast}%
  \BibitemOpen
  \bibfield  {author} {\bibinfo {author} {\bibfnamefont {A.~I.}\ \bibnamefont {Kuleff}}\ and\ \bibinfo {author} {\bibfnamefont {L.~S.}\ \bibnamefont {Cederbaum}},\ }\href@noop {} {\bibfield  {journal} {\bibinfo  {journal} {J. Phys. B: At. Mol. Opt. Phys.}\ }\textbf {\bibinfo {volume} {47}},\ \bibinfo {pages} {124002} (\bibinfo {year} {2014})}\BibitemShut {NoStop}%
\bibitem [{\citenamefont {Despr{\'e}}\ \emph {et~al.}(2015)\citenamefont {Despr{\'e}}, \citenamefont {Marciniak}, \citenamefont {Loriot}, \citenamefont {Galbraith}, \citenamefont {Rouz{\'e}e}, \citenamefont {Vrakking}, \citenamefont {L{\'e}pine},\ and\ \citenamefont {Kuleff}}]{despre2015attosecond}%
  \BibitemOpen
  \bibfield  {author} {\bibinfo {author} {\bibfnamefont {V.}~\bibnamefont {Despr{\'e}}}, \bibinfo {author} {\bibfnamefont {A.}~\bibnamefont {Marciniak}}, \bibinfo {author} {\bibfnamefont {V.}~\bibnamefont {Loriot}}, \bibinfo {author} {\bibfnamefont {M.}~\bibnamefont {Galbraith}}, \bibinfo {author} {\bibfnamefont {A.}~\bibnamefont {Rouz{\'e}e}}, \bibinfo {author} {\bibfnamefont {M.}~\bibnamefont {Vrakking}}, \bibinfo {author} {\bibfnamefont {F.}~\bibnamefont {L{\'e}pine}}, \ and\ \bibinfo {author} {\bibfnamefont {A.}~\bibnamefont {Kuleff}},\ }\href@noop {} {\bibfield  {journal} {\bibinfo  {journal} {J. Phys. Chem. Lett.}\ }\textbf {\bibinfo {volume} {6}},\ \bibinfo {pages} {426} (\bibinfo {year} {2015})}\BibitemShut {NoStop}%
\bibitem [{\citenamefont {Despr{\'e}}\ \emph {et~al.}(2018)\citenamefont {Despr{\'e}}, \citenamefont {Golubev},\ and\ \citenamefont {Kuleff}}]{despre2018charge}%
  \BibitemOpen
  \bibfield  {author} {\bibinfo {author} {\bibfnamefont {V.}~\bibnamefont {Despr{\'e}}}, \bibinfo {author} {\bibfnamefont {N.~V.}\ \bibnamefont {Golubev}}, \ and\ \bibinfo {author} {\bibfnamefont {A.~I.}\ \bibnamefont {Kuleff}},\ }\href@noop {} {\bibfield  {journal} {\bibinfo  {journal} {Phys. Rev. Lett.}\ }\textbf {\bibinfo {volume} {121}},\ \bibinfo {pages} {203002} (\bibinfo {year} {2018})}\BibitemShut {NoStop}%
\bibitem [{\citenamefont {Scheidegger}\ \emph {et~al.}(2022)\citenamefont {Scheidegger}, \citenamefont {Van{\'\i}{\v{c}}ek},\ and\ \citenamefont {Golubev}}]{scheidegger2022search}%
  \BibitemOpen
  \bibfield  {author} {\bibinfo {author} {\bibfnamefont {A.}~\bibnamefont {Scheidegger}}, \bibinfo {author} {\bibfnamefont {J.}~\bibnamefont {Van{\'\i}{\v{c}}ek}}, \ and\ \bibinfo {author} {\bibfnamefont {N.~V.}\ \bibnamefont {Golubev}},\ }\href@noop {} {\bibfield  {journal} {\bibinfo  {journal} {J. Chem. Phys.}\ }\textbf {\bibinfo {volume} {156}},\ \bibinfo {pages} {034104} (\bibinfo {year} {2022})}\BibitemShut {NoStop}%
\bibitem [{\citenamefont {Vacher}\ \emph {et~al.}(2015)\citenamefont {Vacher}, \citenamefont {Steinberg}, \citenamefont {Jenkins}, \citenamefont {Bearpark},\ and\ \citenamefont {Robb}}]{vacher2015electron}%
  \BibitemOpen
  \bibfield  {author} {\bibinfo {author} {\bibfnamefont {M.}~\bibnamefont {Vacher}}, \bibinfo {author} {\bibfnamefont {L.}~\bibnamefont {Steinberg}}, \bibinfo {author} {\bibfnamefont {A.~J.}\ \bibnamefont {Jenkins}}, \bibinfo {author} {\bibfnamefont {M.~J.}\ \bibnamefont {Bearpark}}, \ and\ \bibinfo {author} {\bibfnamefont {M.~A.}\ \bibnamefont {Robb}},\ }\href@noop {} {\bibfield  {journal} {\bibinfo  {journal} {Phys. Rev. A}\ }\textbf {\bibinfo {volume} {92}},\ \bibinfo {pages} {040502} (\bibinfo {year} {2015})}\BibitemShut {NoStop}%
\bibitem [{\citenamefont {Jenkins}\ \emph {et~al.}(2016)\citenamefont {Jenkins}, \citenamefont {Vacher}, \citenamefont {Bearpark},\ and\ \citenamefont {Robb}}]{jenkins2016nuclear}%
  \BibitemOpen
  \bibfield  {author} {\bibinfo {author} {\bibfnamefont {A.~J.}\ \bibnamefont {Jenkins}}, \bibinfo {author} {\bibfnamefont {M.}~\bibnamefont {Vacher}}, \bibinfo {author} {\bibfnamefont {M.~J.}\ \bibnamefont {Bearpark}}, \ and\ \bibinfo {author} {\bibfnamefont {M.~A.}\ \bibnamefont {Robb}},\ }\href@noop {} {\bibfield  {journal} {\bibinfo  {journal} {J. Chem. Phys.}\ }\textbf {\bibinfo {volume} {144}},\ \bibinfo {pages} {104110} (\bibinfo {year} {2016})}\BibitemShut {NoStop}%
\bibitem [{\citenamefont {Vacher}\ \emph {et~al.}(2017)\citenamefont {Vacher}, \citenamefont {Bearpark}, \citenamefont {Robb},\ and\ \citenamefont {Malhado}}]{vacher2017electron}%
  \BibitemOpen
  \bibfield  {author} {\bibinfo {author} {\bibfnamefont {M.}~\bibnamefont {Vacher}}, \bibinfo {author} {\bibfnamefont {M.~J.}\ \bibnamefont {Bearpark}}, \bibinfo {author} {\bibfnamefont {M.~A.}\ \bibnamefont {Robb}}, \ and\ \bibinfo {author} {\bibfnamefont {J.~P.}\ \bibnamefont {Malhado}},\ }\href@noop {} {\bibfield  {journal} {\bibinfo  {journal} {Phys. Rev. Lett.}\ }\textbf {\bibinfo {volume} {118}},\ \bibinfo {pages} {083001} (\bibinfo {year} {2017})}\BibitemShut {NoStop}%
\bibitem [{\citenamefont {Arnold}\ \emph {et~al.}(2017)\citenamefont {Arnold}, \citenamefont {Vendrell},\ and\ \citenamefont {Santra}}]{arnold2017electronic}%
  \BibitemOpen
  \bibfield  {author} {\bibinfo {author} {\bibfnamefont {C.}~\bibnamefont {Arnold}}, \bibinfo {author} {\bibfnamefont {O.}~\bibnamefont {Vendrell}}, \ and\ \bibinfo {author} {\bibfnamefont {R.}~\bibnamefont {Santra}},\ }\href@noop {} {\bibfield  {journal} {\bibinfo  {journal} {Phys. Rev. A}\ }\textbf {\bibinfo {volume} {95}},\ \bibinfo {pages} {033425} (\bibinfo {year} {2017})}\BibitemShut {NoStop}%
\bibitem [{\citenamefont {Kraus}\ \emph {et~al.}(2015)\citenamefont {Kraus}, \citenamefont {Mignolet}, \citenamefont {Baykusheva}, \citenamefont {Rupenyan}, \citenamefont {Horn{\`y}}, \citenamefont {Penka}, \citenamefont {Grassi}, \citenamefont {Tolstikhin}, \citenamefont {Schneider}, \citenamefont {Jensen} \emph {et~al.}}]{kraus2015measurement}%
  \BibitemOpen
  \bibfield  {author} {\bibinfo {author} {\bibfnamefont {P.~M.}\ \bibnamefont {Kraus}}, \bibinfo {author} {\bibfnamefont {B.}~\bibnamefont {Mignolet}}, \bibinfo {author} {\bibfnamefont {D.}~\bibnamefont {Baykusheva}}, \bibinfo {author} {\bibfnamefont {A.}~\bibnamefont {Rupenyan}}, \bibinfo {author} {\bibfnamefont {L.}~\bibnamefont {Horn{\`y}}}, \bibinfo {author} {\bibfnamefont {E.~F.}\ \bibnamefont {Penka}}, \bibinfo {author} {\bibfnamefont {G.}~\bibnamefont {Grassi}}, \bibinfo {author} {\bibfnamefont {O.~I.}\ \bibnamefont {Tolstikhin}}, \bibinfo {author} {\bibfnamefont {J.}~\bibnamefont {Schneider}}, \bibinfo {author} {\bibfnamefont {F.}~\bibnamefont {Jensen}},  \emph {et~al.},\ }\href@noop {} {\bibfield  {journal} {\bibinfo  {journal} {Science}\ }\textbf {\bibinfo {volume} {350}},\ \bibinfo {pages} {790} (\bibinfo {year} {2015})}\BibitemShut {NoStop}%
\bibitem [{\citenamefont {Lara-Astiaso}\ \emph {et~al.}(2018{\natexlab{b}})\citenamefont {Lara-Astiaso}, \citenamefont {Galli}, \citenamefont {Trabattoni}, \citenamefont {Palacios}, \citenamefont {Ayuso}, \citenamefont {Frassetto}, \citenamefont {Poletto}, \citenamefont {De~Camillis}, \citenamefont {Greenwood}, \citenamefont {Decleva} \emph {et~al.}}]{lara2018attosecond}%
  \BibitemOpen
  \bibfield  {author} {\bibinfo {author} {\bibfnamefont {M.}~\bibnamefont {Lara-Astiaso}}, \bibinfo {author} {\bibfnamefont {M.}~\bibnamefont {Galli}}, \bibinfo {author} {\bibfnamefont {A.}~\bibnamefont {Trabattoni}}, \bibinfo {author} {\bibfnamefont {A.}~\bibnamefont {Palacios}}, \bibinfo {author} {\bibfnamefont {D.}~\bibnamefont {Ayuso}}, \bibinfo {author} {\bibfnamefont {F.}~\bibnamefont {Frassetto}}, \bibinfo {author} {\bibfnamefont {L.}~\bibnamefont {Poletto}}, \bibinfo {author} {\bibfnamefont {S.}~\bibnamefont {De~Camillis}}, \bibinfo {author} {\bibfnamefont {J.}~\bibnamefont {Greenwood}}, \bibinfo {author} {\bibfnamefont {P.}~\bibnamefont {Decleva}},  \emph {et~al.},\ }\href@noop {} {\bibfield  {journal} {\bibinfo  {journal} {J. Phys. Chem. Lett.}\ }\textbf {\bibinfo {volume} {9}},\ \bibinfo {pages} {4570} (\bibinfo {year} {2018}{\natexlab{b}})}\BibitemShut {NoStop}%
\bibitem [{\citenamefont {Matselyukh}\ \emph {et~al.}(2022)\citenamefont {Matselyukh}, \citenamefont {Despr{\'e}}, \citenamefont {Golubev}, \citenamefont {Kuleff},\ and\ \citenamefont {W{\"o}rner}}]{matselyukh2022decoherence}%
  \BibitemOpen
  \bibfield  {author} {\bibinfo {author} {\bibfnamefont {D.~T.}\ \bibnamefont {Matselyukh}}, \bibinfo {author} {\bibfnamefont {V.}~\bibnamefont {Despr{\'e}}}, \bibinfo {author} {\bibfnamefont {N.~V.}\ \bibnamefont {Golubev}}, \bibinfo {author} {\bibfnamefont {A.~I.}\ \bibnamefont {Kuleff}}, \ and\ \bibinfo {author} {\bibfnamefont {H.~J.}\ \bibnamefont {W{\"o}rner}},\ }\href@noop {} {\bibfield  {journal} {\bibinfo  {journal} {Nat. Phys.}\ }\textbf {\bibinfo {volume} {18}},\ \bibinfo {pages} {1206} (\bibinfo {year} {2022})}\BibitemShut {NoStop}%
\bibitem [{\citenamefont {Jia}\ \emph {et~al.}(2019)\citenamefont {Jia}, \citenamefont {Manz},\ and\ \citenamefont {Yang}}]{Jia2019timing}%
  \BibitemOpen
  \bibfield  {author} {\bibinfo {author} {\bibfnamefont {D.}~\bibnamefont {Jia}}, \bibinfo {author} {\bibfnamefont {J.}~\bibnamefont {Manz}}, \ and\ \bibinfo {author} {\bibfnamefont {Y.}~\bibnamefont {Yang}},\ }\href@noop {} {\bibfield  {journal} {\bibinfo  {journal} {J. Chem. Phys.}\ }\textbf {\bibinfo {volume} {151}},\ \bibinfo {pages} {244306} (\bibinfo {year} {2019})}\BibitemShut {NoStop}%
\bibitem [{\citenamefont {M{\aa}nsson}\ \emph {et~al.}(2021)\citenamefont {M{\aa}nsson}, \citenamefont {Latini}, \citenamefont {Covito}, \citenamefont {Wanie}, \citenamefont {Galli}, \citenamefont {Perfetto}, \citenamefont {Stefanucci}, \citenamefont {H{\"u}bener}, \citenamefont {De~Giovannini}, \citenamefont {Castrovilli} \emph {et~al.}}]{maansson2021real}%
  \BibitemOpen
  \bibfield  {author} {\bibinfo {author} {\bibfnamefont {E.~P.}\ \bibnamefont {M{\aa}nsson}}, \bibinfo {author} {\bibfnamefont {S.}~\bibnamefont {Latini}}, \bibinfo {author} {\bibfnamefont {F.}~\bibnamefont {Covito}}, \bibinfo {author} {\bibfnamefont {V.}~\bibnamefont {Wanie}}, \bibinfo {author} {\bibfnamefont {M.}~\bibnamefont {Galli}}, \bibinfo {author} {\bibfnamefont {E.}~\bibnamefont {Perfetto}}, \bibinfo {author} {\bibfnamefont {G.}~\bibnamefont {Stefanucci}}, \bibinfo {author} {\bibfnamefont {H.}~\bibnamefont {H{\"u}bener}}, \bibinfo {author} {\bibfnamefont {U.}~\bibnamefont {De~Giovannini}}, \bibinfo {author} {\bibfnamefont {M.~C.}\ \bibnamefont {Castrovilli}},  \emph {et~al.},\ }\href@noop {} {\bibfield  {journal} {\bibinfo  {journal} {Commun. Chem.}\ }\textbf {\bibinfo {volume} {4}},\ \bibinfo {pages} {73} (\bibinfo {year} {2021})}\BibitemShut {NoStop}%
\bibitem [{\citenamefont {Despr{\'e}}\ and\ \citenamefont {Kuleff}(2022)}]{despre2022correlation}%
  \BibitemOpen
  \bibfield  {author} {\bibinfo {author} {\bibfnamefont {V.}~\bibnamefont {Despr{\'e}}}\ and\ \bibinfo {author} {\bibfnamefont {A.~I.}\ \bibnamefont {Kuleff}},\ }\href@noop {} {\bibfield  {journal} {\bibinfo  {journal} {Phys. Rev. A}\ }\textbf {\bibinfo {volume} {106}},\ \bibinfo {pages} {L021501} (\bibinfo {year} {2022})}\BibitemShut {NoStop}%
\bibitem [{\citenamefont {Cederbaum}\ \emph {et~al.}(1986)\citenamefont {Cederbaum}, \citenamefont {Domcke}, \citenamefont {Schirmer},\ and\ \citenamefont {Niessen}}]{cederbaum1986correlation}%
  \BibitemOpen
  \bibfield  {author} {\bibinfo {author} {\bibfnamefont {L.~S.}\ \bibnamefont {Cederbaum}}, \bibinfo {author} {\bibfnamefont {W.}~\bibnamefont {Domcke}}, \bibinfo {author} {\bibfnamefont {J.}~\bibnamefont {Schirmer}}, \ and\ \bibinfo {author} {\bibfnamefont {W.~v.}\ \bibnamefont {Niessen}},\ }\href@noop {} {\bibfield  {journal} {\bibinfo  {journal} {Adv. Chem. Phys.}\ ,\ \bibinfo {pages} {115}} (\bibinfo {year} {1986})}\BibitemShut {NoStop}%
\bibitem [{\citenamefont {Deleuze}\ and\ \citenamefont {Cederbaum}(1996)}]{deleuze1996formation}%
  \BibitemOpen
  \bibfield  {author} {\bibinfo {author} {\bibfnamefont {M.~S.}\ \bibnamefont {Deleuze}}\ and\ \bibinfo {author} {\bibfnamefont {L.~S.}\ \bibnamefont {Cederbaum}},\ }\href@noop {} {\bibfield  {journal} {\bibinfo  {journal} {Phys. Rev. B}\ }\textbf {\bibinfo {volume} {53}},\ \bibinfo {pages} {13326} (\bibinfo {year} {1996})}\BibitemShut {NoStop}%
\bibitem [{\citenamefont {Herv{\'e}}\ \emph {et~al.}(2021)\citenamefont {Herv{\'e}}, \citenamefont {Despr{\'e}}, \citenamefont {Castellanos~Nash}, \citenamefont {Loriot}, \citenamefont {Boyer}, \citenamefont {Scognamiglio}, \citenamefont {Karras}, \citenamefont {Br{\'e}dy}, \citenamefont {Constant}, \citenamefont {Tielens} \emph {et~al.}}]{herve2021ultrafast}%
  \BibitemOpen
  \bibfield  {author} {\bibinfo {author} {\bibfnamefont {M.}~\bibnamefont {Herv{\'e}}}, \bibinfo {author} {\bibfnamefont {V.}~\bibnamefont {Despr{\'e}}}, \bibinfo {author} {\bibfnamefont {P.}~\bibnamefont {Castellanos~Nash}}, \bibinfo {author} {\bibfnamefont {V.}~\bibnamefont {Loriot}}, \bibinfo {author} {\bibfnamefont {A.}~\bibnamefont {Boyer}}, \bibinfo {author} {\bibfnamefont {A.}~\bibnamefont {Scognamiglio}}, \bibinfo {author} {\bibfnamefont {G.}~\bibnamefont {Karras}}, \bibinfo {author} {\bibfnamefont {R.}~\bibnamefont {Br{\'e}dy}}, \bibinfo {author} {\bibfnamefont {E.}~\bibnamefont {Constant}}, \bibinfo {author} {\bibfnamefont {A.}~\bibnamefont {Tielens}},  \emph {et~al.},\ }\href@noop {} {\bibfield  {journal} {\bibinfo  {journal} {Nat. Phys.}\ }\textbf {\bibinfo {volume} {17}},\ \bibinfo {pages} {327} (\bibinfo {year} {2021})}\BibitemShut {NoStop}%
\bibitem [{\citenamefont {Belshaw}\ \emph {et~al.}(2012)\citenamefont {Belshaw}, \citenamefont {Calegari}, \citenamefont {Duffy}, \citenamefont {Trabattoni}, \citenamefont {Poletto}, \citenamefont {Nisoli},\ and\ \citenamefont {Greenwood}}]{belshaw2012observation}%
  \BibitemOpen
  \bibfield  {author} {\bibinfo {author} {\bibfnamefont {L.}~\bibnamefont {Belshaw}}, \bibinfo {author} {\bibfnamefont {F.}~\bibnamefont {Calegari}}, \bibinfo {author} {\bibfnamefont {M.~J.}\ \bibnamefont {Duffy}}, \bibinfo {author} {\bibfnamefont {A.}~\bibnamefont {Trabattoni}}, \bibinfo {author} {\bibfnamefont {L.}~\bibnamefont {Poletto}}, \bibinfo {author} {\bibfnamefont {M.}~\bibnamefont {Nisoli}}, \ and\ \bibinfo {author} {\bibfnamefont {J.~B.}\ \bibnamefont {Greenwood}},\ }\href@noop {} {\bibfield  {journal} {\bibinfo  {journal} {J. Phys. Chem. Lett.}\ }\textbf {\bibinfo {volume} {3}},\ \bibinfo {pages} {3751} (\bibinfo {year} {2012})}\BibitemShut {NoStop}%
\bibitem [{\citenamefont {Marciniak}\ \emph {et~al.}(2015)\citenamefont {Marciniak}, \citenamefont {Despr{\'e}}, \citenamefont {Barillot}, \citenamefont {Rouz{\'e}e}, \citenamefont {Galbraith}, \citenamefont {Klei}, \citenamefont {Yang}, \citenamefont {Smeenk}, \citenamefont {Loriot}, \citenamefont {Reddy} \emph {et~al.}}]{marciniak2015xuv}%
  \BibitemOpen
  \bibfield  {author} {\bibinfo {author} {\bibfnamefont {A.}~\bibnamefont {Marciniak}}, \bibinfo {author} {\bibfnamefont {V.}~\bibnamefont {Despr{\'e}}}, \bibinfo {author} {\bibfnamefont {T.}~\bibnamefont {Barillot}}, \bibinfo {author} {\bibfnamefont {A.}~\bibnamefont {Rouz{\'e}e}}, \bibinfo {author} {\bibfnamefont {M.}~\bibnamefont {Galbraith}}, \bibinfo {author} {\bibfnamefont {J.}~\bibnamefont {Klei}}, \bibinfo {author} {\bibfnamefont {C.-H.}\ \bibnamefont {Yang}}, \bibinfo {author} {\bibfnamefont {C.}~\bibnamefont {Smeenk}}, \bibinfo {author} {\bibfnamefont {V.}~\bibnamefont {Loriot}}, \bibinfo {author} {\bibfnamefont {S.~N.}\ \bibnamefont {Reddy}},  \emph {et~al.},\ }\href@noop {} {\bibfield  {journal} {\bibinfo  {journal} {Nat. Commun.}\ }\textbf {\bibinfo {volume} {6}},\ \bibinfo {pages} {7909} (\bibinfo {year} {2015})}\BibitemShut {NoStop}%
\bibitem [{\citenamefont {Marciniak}\ \emph {et~al.}(2018)\citenamefont {Marciniak}, \citenamefont {Yamazaki}, \citenamefont {Maeda}, \citenamefont {Reduzzi}, \citenamefont {Despr{\'e}}, \citenamefont {Herv{\'e}}, \citenamefont {Meziane}, \citenamefont {Niehaus}, \citenamefont {Loriot}, \citenamefont {Kuleff} \emph {et~al.}}]{marciniak2018ultrafast}%
  \BibitemOpen
  \bibfield  {author} {\bibinfo {author} {\bibfnamefont {A.}~\bibnamefont {Marciniak}}, \bibinfo {author} {\bibfnamefont {K.}~\bibnamefont {Yamazaki}}, \bibinfo {author} {\bibfnamefont {S.}~\bibnamefont {Maeda}}, \bibinfo {author} {\bibfnamefont {M.}~\bibnamefont {Reduzzi}}, \bibinfo {author} {\bibfnamefont {V.}~\bibnamefont {Despr{\'e}}}, \bibinfo {author} {\bibfnamefont {M.}~\bibnamefont {Herv{\'e}}}, \bibinfo {author} {\bibfnamefont {M.}~\bibnamefont {Meziane}}, \bibinfo {author} {\bibfnamefont {T.~A.}\ \bibnamefont {Niehaus}}, \bibinfo {author} {\bibfnamefont {V.}~\bibnamefont {Loriot}}, \bibinfo {author} {\bibfnamefont {A.~I.}\ \bibnamefont {Kuleff}},  \emph {et~al.},\ }\href@noop {} {\bibfield  {journal} {\bibinfo  {journal} {J. Phys. Chem. Lett.}\ }\textbf {\bibinfo {volume} {9}},\ \bibinfo {pages} {6927} (\bibinfo {year} {2018})}\BibitemShut {NoStop}%
\bibitem [{\citenamefont {Boyer}\ \emph {et~al.}(2021)\citenamefont {Boyer}, \citenamefont {Herv{\'e}}, \citenamefont {Despr{\'e}}, \citenamefont {Nash}, \citenamefont {Loriot}, \citenamefont {Marciniak}, \citenamefont {Tielens}, \citenamefont {Kuleff},\ and\ \citenamefont {L{\'e}pine}}]{boyer2021ultrafast}%
  \BibitemOpen
  \bibfield  {author} {\bibinfo {author} {\bibfnamefont {A.}~\bibnamefont {Boyer}}, \bibinfo {author} {\bibfnamefont {M.}~\bibnamefont {Herv{\'e}}}, \bibinfo {author} {\bibfnamefont {V.}~\bibnamefont {Despr{\'e}}}, \bibinfo {author} {\bibfnamefont {P.~C.}\ \bibnamefont {Nash}}, \bibinfo {author} {\bibfnamefont {V.}~\bibnamefont {Loriot}}, \bibinfo {author} {\bibfnamefont {A.}~\bibnamefont {Marciniak}}, \bibinfo {author} {\bibfnamefont {A.}~\bibnamefont {Tielens}}, \bibinfo {author} {\bibfnamefont {A.}~\bibnamefont {Kuleff}}, \ and\ \bibinfo {author} {\bibfnamefont {F.}~\bibnamefont {L{\'e}pine}},\ }\href@noop {} {\bibfield  {journal} {\bibinfo  {journal} {Phys. Rev. X}\ }\textbf {\bibinfo {volume} {11}},\ \bibinfo {pages} {041012} (\bibinfo {year} {2021})}\BibitemShut {NoStop}%
\bibitem [{\citenamefont {Despr{\'e}}\ and\ \citenamefont {Kuleff}(2019)}]{despre2019size}%
  \BibitemOpen
  \bibfield  {author} {\bibinfo {author} {\bibfnamefont {V.}~\bibnamefont {Despr{\'e}}}\ and\ \bibinfo {author} {\bibfnamefont {A.~I.}\ \bibnamefont {Kuleff}},\ }\href@noop {} {\bibfield  {journal} {\bibinfo  {journal} {Theor. Chem. Acc.}\ }\textbf {\bibinfo {volume} {138}},\ \bibinfo {pages} {1} (\bibinfo {year} {2019})}\BibitemShut {NoStop}%
\bibitem [{\citenamefont {Folorunso}\ \emph {et~al.}(2021)\citenamefont {Folorunso}, \citenamefont {Bruner}, \citenamefont {Mauger}, \citenamefont {Hamer}, \citenamefont {Hernandez}, \citenamefont {Jones}, \citenamefont {DiMauro}, \citenamefont {Gaarde}, \citenamefont {Schafer},\ and\ \citenamefont {Lopata}}]{folorunso2021molecular}%
  \BibitemOpen
  \bibfield  {author} {\bibinfo {author} {\bibfnamefont {A.~S.}\ \bibnamefont {Folorunso}}, \bibinfo {author} {\bibfnamefont {A.}~\bibnamefont {Bruner}}, \bibinfo {author} {\bibfnamefont {F.}~\bibnamefont {Mauger}}, \bibinfo {author} {\bibfnamefont {K.~A.}\ \bibnamefont {Hamer}}, \bibinfo {author} {\bibfnamefont {S.}~\bibnamefont {Hernandez}}, \bibinfo {author} {\bibfnamefont {R.~R.}\ \bibnamefont {Jones}}, \bibinfo {author} {\bibfnamefont {L.~F.}\ \bibnamefont {DiMauro}}, \bibinfo {author} {\bibfnamefont {M.~B.}\ \bibnamefont {Gaarde}}, \bibinfo {author} {\bibfnamefont {K.~J.}\ \bibnamefont {Schafer}}, \ and\ \bibinfo {author} {\bibfnamefont {K.}~\bibnamefont {Lopata}},\ }\href@noop {} {\bibfield  {journal} {\bibinfo  {journal} {Phys. Rev. Lett.}\ }\textbf {\bibinfo {volume} {126}},\ \bibinfo {pages} {133002} (\bibinfo {year} {2021})}\BibitemShut {NoStop}%
\bibitem [{\citenamefont {Mauger}\ \emph {et~al.}(2022)\citenamefont {Mauger}, \citenamefont {Folorunso}, \citenamefont {Hamer}, \citenamefont {Chandre}, \citenamefont {Gaarde}, \citenamefont {Lopata},\ and\ \citenamefont {Schafer}}]{mauger2022charge}%
  \BibitemOpen
  \bibfield  {author} {\bibinfo {author} {\bibfnamefont {F.}~\bibnamefont {Mauger}}, \bibinfo {author} {\bibfnamefont {A.~S.}\ \bibnamefont {Folorunso}}, \bibinfo {author} {\bibfnamefont {K.~A.}\ \bibnamefont {Hamer}}, \bibinfo {author} {\bibfnamefont {C.}~\bibnamefont {Chandre}}, \bibinfo {author} {\bibfnamefont {M.~B.}\ \bibnamefont {Gaarde}}, \bibinfo {author} {\bibfnamefont {K.}~\bibnamefont {Lopata}}, \ and\ \bibinfo {author} {\bibfnamefont {K.~J.}\ \bibnamefont {Schafer}},\ }\href@noop {} {\bibfield  {journal} {\bibinfo  {journal} {Phys. Rev. Res.}\ }\textbf {\bibinfo {volume} {4}},\ \bibinfo {pages} {013073} (\bibinfo {year} {2022})}\BibitemShut {NoStop}%
\bibitem [{\citenamefont {Chordiya}\ \emph {et~al.}(2023)\citenamefont {Chordiya}, \citenamefont {Despr{\'e}}, \citenamefont {Nagyill{\'e}s}, \citenamefont {Zeller}, \citenamefont {Diveki}, \citenamefont {Kuleff},\ and\ \citenamefont {Kahaly}}]{chordiya2023photo}%
  \BibitemOpen
  \bibfield  {author} {\bibinfo {author} {\bibfnamefont {K.}~\bibnamefont {Chordiya}}, \bibinfo {author} {\bibfnamefont {V.}~\bibnamefont {Despr{\'e}}}, \bibinfo {author} {\bibfnamefont {B.}~\bibnamefont {Nagyill{\'e}s}}, \bibinfo {author} {\bibfnamefont {F.}~\bibnamefont {Zeller}}, \bibinfo {author} {\bibfnamefont {Z.}~\bibnamefont {Diveki}}, \bibinfo {author} {\bibfnamefont {A.~I.}\ \bibnamefont {Kuleff}}, \ and\ \bibinfo {author} {\bibfnamefont {M.~U.}\ \bibnamefont {Kahaly}},\ }\href@noop {} {\bibfield  {journal} {\bibinfo  {journal} {Phys. Chem. Chem. Phys.}\ }\textbf {\bibinfo {volume} {25}},\ \bibinfo {pages} {4472} (\bibinfo {year} {2023})}\BibitemShut {NoStop}%
\bibitem [{\citenamefont {Xiang}\ \emph {et~al.}(2016)\citenamefont {Xiang}, \citenamefont {Wang}, \citenamefont {Jia}, \citenamefont {Lee},\ and\ \citenamefont {Guo}}]{xiang2016molecular}%
  \BibitemOpen
  \bibfield  {author} {\bibinfo {author} {\bibfnamefont {D.}~\bibnamefont {Xiang}}, \bibinfo {author} {\bibfnamefont {X.}~\bibnamefont {Wang}}, \bibinfo {author} {\bibfnamefont {C.}~\bibnamefont {Jia}}, \bibinfo {author} {\bibfnamefont {T.}~\bibnamefont {Lee}}, \ and\ \bibinfo {author} {\bibfnamefont {X.}~\bibnamefont {Guo}},\ }\href@noop {} {\bibfield  {journal} {\bibinfo  {journal} {Chem. Rev.}\ }\textbf {\bibinfo {volume} {116}},\ \bibinfo {pages} {4318} (\bibinfo {year} {2016})}\BibitemShut {NoStop}%
\bibitem [{\citenamefont {Harriman}\ and\ \citenamefont {Ziessel}(1996)}]{harriman1996making}%
  \BibitemOpen
  \bibfield  {author} {\bibinfo {author} {\bibfnamefont {A.}~\bibnamefont {Harriman}}\ and\ \bibinfo {author} {\bibfnamefont {R.}~\bibnamefont {Ziessel}},\ }\href@noop {} {\bibfield  {journal} {\bibinfo  {journal} {Chem. Comm.}\ ,\ \bibinfo {pages} {1707}} (\bibinfo {year} {1996})}\BibitemShut {NoStop}%
\bibitem [{\citenamefont {Grosshenny}\ \emph {et~al.}(1996)\citenamefont {Grosshenny}, \citenamefont {Harriman},\ and\ \citenamefont {Ziessel}}]{grosshenny1996towards}%
  \BibitemOpen
  \bibfield  {author} {\bibinfo {author} {\bibfnamefont {V.}~\bibnamefont {Grosshenny}}, \bibinfo {author} {\bibfnamefont {A.}~\bibnamefont {Harriman}}, \ and\ \bibinfo {author} {\bibfnamefont {R.}~\bibnamefont {Ziessel}},\ }\href@noop {} {\bibfield  {journal} {\bibinfo  {journal} {Angew. Chem., Int. Ed. Engl.}\ }\textbf {\bibinfo {volume} {34}},\ \bibinfo {pages} {2705} (\bibinfo {year} {1996})}\BibitemShut {NoStop}%
\bibitem [{\citenamefont {James}\ and\ \citenamefont {Tour}(2005)}]{james2005molecular}%
  \BibitemOpen
  \bibfield  {author} {\bibinfo {author} {\bibfnamefont {D.~K.}\ \bibnamefont {James}}\ and\ \bibinfo {author} {\bibfnamefont {J.~M.}\ \bibnamefont {Tour}},\ }in\ \href@noop {} {\emph {\bibinfo {booktitle} {Molecular Wires and Electronics}}}\ (\bibinfo  {publisher} {Springer},\ \bibinfo {year} {2005})\ pp.\ \bibinfo {pages} {33--62}\BibitemShut {NoStop}%
\bibitem [{\citenamefont {Schirmer}(1982)}]{schirmer1982beyond}%
  \BibitemOpen
  \bibfield  {author} {\bibinfo {author} {\bibfnamefont {J.}~\bibnamefont {Schirmer}},\ }\href@noop {} {\bibfield  {journal} {\bibinfo  {journal} {Phys. Rev. A}\ }\textbf {\bibinfo {volume} {26}},\ \bibinfo {pages} {2395} (\bibinfo {year} {1982})}\BibitemShut {NoStop}%
\bibitem [{\citenamefont {Schirmer}(2018)}]{schirmer2018many}%
  \BibitemOpen
  \bibfield  {author} {\bibinfo {author} {\bibfnamefont {J.}~\bibnamefont {Schirmer}},\ }\href@noop {} {\emph {\bibinfo {title} {Many-body methods for atoms, molecules and clusters}}}\ (\bibinfo  {publisher} {Springer},\ \bibinfo {year} {2018})\BibitemShut {NoStop}%
\bibitem [{\citenamefont {Schirmer}\ \emph {et~al.}(1998)\citenamefont {Schirmer}, \citenamefont {Trofimov},\ and\ \citenamefont {Stelter}}]{schirmer1998non}%
  \BibitemOpen
  \bibfield  {author} {\bibinfo {author} {\bibfnamefont {J.}~\bibnamefont {Schirmer}}, \bibinfo {author} {\bibfnamefont {A.~B.}\ \bibnamefont {Trofimov}}, \ and\ \bibinfo {author} {\bibfnamefont {G.}~\bibnamefont {Stelter}},\ }\href@noop {} {\bibfield  {journal} {\bibinfo  {journal} {J. Chem. Phys.}\ }\textbf {\bibinfo {volume} {109}},\ \bibinfo {pages} {4734} (\bibinfo {year} {1998})}\BibitemShut {NoStop}%
\bibitem [{\citenamefont {Frisch}\ \emph {et~al.}(2009)\citenamefont {Frisch}, \citenamefont {Trucks}, \citenamefont {Schlegel}, \citenamefont {Scuseria}, \citenamefont {Robb}, \citenamefont {Cheeseman}, \citenamefont {Scalmani}, \citenamefont {Barone}, \citenamefont {Mennucci}, \citenamefont {Petersson} \emph {et~al.}}]{frisch2009uranyl}%
  \BibitemOpen
  \bibfield  {author} {\bibinfo {author} {\bibfnamefont {M.}~\bibnamefont {Frisch}}, \bibinfo {author} {\bibfnamefont {G.}~\bibnamefont {Trucks}}, \bibinfo {author} {\bibfnamefont {H.}~\bibnamefont {Schlegel}}, \bibinfo {author} {\bibfnamefont {G.}~\bibnamefont {Scuseria}}, \bibinfo {author} {\bibfnamefont {M.}~\bibnamefont {Robb}}, \bibinfo {author} {\bibfnamefont {J.}~\bibnamefont {Cheeseman}}, \bibinfo {author} {\bibfnamefont {G.}~\bibnamefont {Scalmani}}, \bibinfo {author} {\bibfnamefont {V.}~\bibnamefont {Barone}}, \bibinfo {author} {\bibfnamefont {B.}~\bibnamefont {Mennucci}}, \bibinfo {author} {\bibfnamefont {G.}~\bibnamefont {Petersson}},  \emph {et~al.},\ }\bibfield  {booktitle} {\emph {\bibinfo {booktitle} {Gaussian 09}},\ }\href@noop {} {\  (\bibinfo {year} {2009})}\BibitemShut {NoStop}%
\bibitem [{\citenamefont {Schirmer}\ and\ \citenamefont {Barth}(1984)}]{schirmer1984higher}%
  \BibitemOpen
  \bibfield  {author} {\bibinfo {author} {\bibfnamefont {J.}~\bibnamefont {Schirmer}}\ and\ \bibinfo {author} {\bibfnamefont {A.}~\bibnamefont {Barth}},\ }\href@noop {} {\bibfield  {journal} {\bibinfo  {journal} {Z. Phys. A}\ }\textbf {\bibinfo {volume} {317}},\ \bibinfo {pages} {267} (\bibinfo {year} {1984})}\BibitemShut {NoStop}%
\bibitem [{\citenamefont {Kuleff}\ \emph {et~al.}(2005)\citenamefont {Kuleff}, \citenamefont {Breidbach},\ and\ \citenamefont {Cederbaum}}]{kuleff2005multielectron}%
  \BibitemOpen
  \bibfield  {author} {\bibinfo {author} {\bibfnamefont {A.~I.}\ \bibnamefont {Kuleff}}, \bibinfo {author} {\bibfnamefont {J.}~\bibnamefont {Breidbach}}, \ and\ \bibinfo {author} {\bibfnamefont {L.~S.}\ \bibnamefont {Cederbaum}},\ }\href@noop {} {\bibfield  {journal} {\bibinfo  {journal} {J. Chem. Phys.}\ }\textbf {\bibinfo {volume} {123}} (\bibinfo {year} {2005})}\BibitemShut {NoStop}%
\bibitem [{\citenamefont {Breidbach}\ and\ \citenamefont {Cederbaum}(2003)}]{breidbach2003migration}%
  \BibitemOpen
  \bibfield  {author} {\bibinfo {author} {\bibfnamefont {J.}~\bibnamefont {Breidbach}}\ and\ \bibinfo {author} {\bibfnamefont {L.}~\bibnamefont {Cederbaum}},\ }\href@noop {} {\bibfield  {journal} {\bibinfo  {journal} {J. Chem. Phys.}\ }\textbf {\bibinfo {volume} {118}},\ \bibinfo {pages} {3983} (\bibinfo {year} {2003})}\BibitemShut {NoStop}%
\bibitem [{\citenamefont {Kuleff}(2018)}]{kuleff2018ultrafast}%
  \BibitemOpen
  \bibfield  {author} {\bibinfo {author} {\bibfnamefont {A.~I.}\ \bibnamefont {Kuleff}},\ }in\ \href@noop {} {\emph {\bibinfo {booktitle} {Attosecond Molecular Dynamics}}},\ \bibinfo {editor} {edited by\ \bibinfo {editor} {\bibfnamefont {M.~J.~J.}\ \bibnamefont {Vrakking}}\ and\ \bibinfo {editor} {\bibfnamefont {L.}~\bibnamefont {Franck}}}\ (\bibinfo  {publisher} {The Royal Society of Chemistry},\ \bibinfo {year} {2018})\ pp.\ \bibinfo {pages} {103--138}\BibitemShut {NoStop}%
\bibitem [{\citenamefont {Kuleff}\ and\ \citenamefont {Cederbaum}(2007)}]{Kuleff2007Tracing}%
  \BibitemOpen
  \bibfield  {author} {\bibinfo {author} {\bibfnamefont {A.~I.}\ \bibnamefont {Kuleff}}\ and\ \bibinfo {author} {\bibfnamefont {L.~S.}\ \bibnamefont {Cederbaum}},\ }\href@noop {} {\bibfield  {journal} {\bibinfo  {journal} {Phys. Rev. Lett.}\ }\textbf {\bibinfo {volume} {98}},\ \bibinfo {pages} {083201} (\bibinfo {year} {2007})}\BibitemShut {NoStop}%
\bibitem [{\citenamefont {Kuleff}(2017)}]{kuleff2017electronic}%
  \BibitemOpen
  \bibfield  {author} {\bibinfo {author} {\bibfnamefont {A.~I.}\ \bibnamefont {Kuleff}},\ }\href@noop {} {\bibfield  {journal} {\bibinfo  {journal} {Chem. Phys.}\ }\textbf {\bibinfo {volume} {482}},\ \bibinfo {pages} {216} (\bibinfo {year} {2017})}\BibitemShut {NoStop}%
\bibitem [{\citenamefont {Mullenix}\ \emph {et~al.}(2020)\citenamefont {Mullenix}, \citenamefont {Despr{\'{e}}},\ and\ \citenamefont {Kuleff}}]{Mullenix2020Electronic}%
  \BibitemOpen
  \bibfield  {author} {\bibinfo {author} {\bibfnamefont {J.~B.}\ \bibnamefont {Mullenix}}, \bibinfo {author} {\bibfnamefont {V.}~\bibnamefont {Despr{\'{e}}}}, \ and\ \bibinfo {author} {\bibfnamefont {A.~I.}\ \bibnamefont {Kuleff}},\ }\href@noop {} {\bibfield  {journal} {\bibinfo  {journal} {J. Phys. B: At. Mol. Opt. Phys.}\ }\textbf {\bibinfo {volume} {53}},\ \bibinfo {pages} {184006} (\bibinfo {year} {2020})}\BibitemShut {NoStop}%
\bibitem [{\citenamefont {Golubev}\ and\ \citenamefont {Kuleff}(2015)}]{golubev2015control}%
  \BibitemOpen
  \bibfield  {author} {\bibinfo {author} {\bibfnamefont {N.~V.}\ \bibnamefont {Golubev}}\ and\ \bibinfo {author} {\bibfnamefont {A.~I.}\ \bibnamefont {Kuleff}},\ }\href@noop {} {\bibfield  {journal} {\bibinfo  {journal} {Phys. Rev. A}\ }\textbf {\bibinfo {volume} {91}},\ \bibinfo {pages} {051401} (\bibinfo {year} {2015})}\BibitemShut {NoStop}%
\end{thebibliography}%

\end{document}